\documentclass[floatfix,aps,amsmath,nofootinbib,onecolumn,10pt]{revtex4}

\usepackage{subfigure}
\usepackage{graphicx}
\usepackage{bm}
\usepackage{rotating}
\usepackage{array}

\def\({\left(}
\def\){\right)}
\def\[{\left[}
\def\]{\right]}

\def\e{\begin{equation}}
\def\q{\end{equation}}
\def\m{\begin{eqnarray}}
\def\n{\end{eqnarray}}

\begin{document}

\title{Consistency test of the fine-structure constant from the whole ionization history}

\author{Ke Wang$^{1}$ \footnote{wangkey@lzu.edu.cn} and Lu Chen$^{2}$ \footnote{chenlu@sdnu.edu.cn}$^,$ \footnote{Corresponding author}}
\affiliation{$^1$ Lanzhou Center for Theoretical Physics, Key Laboratory of Theoretical Physics of Gansu Province, School of Physical Science and Technology, Lanzhou University, Lanzhou 730000, China\\
Institute of Theoretical Physics \& Research Center of Gravitation, Lanzhou University, Lanzhou 730000, China\\
$^2$ School of Physics and Electronics, \\Shandong Normal University, Jinan 250014, China\\} 
\date{\today}

\begin{abstract}

In cosmology, the fine-structure constant can affect the whole ionization history. However, the previous works confine themselves to the recombination epoch and give various strong constraints on the fine-structure constant. In this paper, we also take the reionization epoch into consideration and do a consistency test of the fine-structure constant from the whole ionization history. 
From the data combination of Planck 2018, BAO data, SNIa samples, SFR density from UV and IR measurements, and the $Q_\text{HII}$ constraints, we find the constraint on the fine-structure constant during the recombination epoch is $\alpha_{\text{rec}}/\alpha_{\text{EM}}=1.001494^{+0.002041}_{-0.002063}$ and its counterpart during the reionization epoch is $\alpha_{\text{rei}}/\alpha_{\text{EM}}=0.854034^{+0.031678}_{-0.027209}$ at 68$\%$ C.L.. They are not consistent with each other by $4.64\sigma$. 
A conservative explanation for such a discrepancy is that there are some issues in the data we used.
We prefer a calibration of some important parameters involved in reconstructing the reionization history.  

\end{abstract}

\pacs{???}

\maketitle


\section{Introduction} 
\label{sec:int}

The fine-structure constant $\alpha_{\text{EM}} =e^2/4\pi \epsilon_0 \hbar c$ is a fundamental constant in physics, which indicates the strength of electromagnetic interaction. 
Its value has been measured with various experiments in laboratory, such as the measurements of the neutron de Broglie wavelength, the quantum Hall effect, the electron anomalous magnetic moment, local atomic clock measurements and so forth~\cite{Kinoshita:1996vz,Lange:2020cul}.
The full 2018 CODATA Committee on Data for Science and Technology~\cite{$https://physics.nist.gov/$} gives the numerical value of $\alpha_{\text{EM}}=1/137.035999084(21)$, its relative standard uncertainty is $1.5\times 10^{-10}$. 
In astrophysical and cosmological systems, we also can get relatively precise measurements of the fine-structure constant with high-resolution astrophysical spectroscopy measurements~\cite{Webb:2010hc,Murphy:2017xaz,Wilczynska:2020rxx,Milakovic:2020tvq,Murphy:2021xhb,daFonseca:2022qdf}, big bang nucleosynthesis~\cite{Deal:2021kjs} and cosmic microwave background (CMB)~\cite{Planck:2014ylh,Hart:2017ndk,Hart:2019dxi,Hart:2021kad} respectively.
For reviews of the other measurements, see~\cite{Uzan:2010pm}.
Since these measurements may come from different epochs of our universe, it is possible for people to probe a varying fine-structure constant \cite{Hart:2021kad,daFonseca:2022qdf,Liu:2021mfk}.
Theoretically, the fine-structure constant can be dynamical in a more general framework than the standard model of particle physics. If it is actually a dynamical quantity, the basic equations related to $\alpha_{\rm EM}$ we are using are approximations of some more general equations.
In practice, comparing all above measurements, one can find that these measurements are consistent with each other and there is no evidence to support a varying fine-structure constant now.

However, there is a caveat for measurements with CMB~\cite{Planck:2014ylh,Hart:2017ndk,Hart:2019dxi,Hart:2021kad} that these previous constraints on the fine-structure constant are only obtained from recombination epoch although the fine-structure constant can affect the whole ionization history.
For recombination epoch, the story is simple. The fine-structure constant influences the Thomson scattering cross-section and the energy levels of hydrogen and helium directly. 
As a result, its derived effects on the visibility function can leave footprints on the CMB power spectra, which can be researched with cosmological observations.
As for the reionization epoch, the situation is complicated. Because there is a lack of knowledge about the reionization, such as its sources and its outset.
Here we suppose that the star-forming galaxies are the major sources of the reionization. Therefore, there are two important parameters for reconstruting the reionization history: the ionizing photon emission rate of a star-forming galaxy $\xi_{\rm ion}$ and the escape fraction of ionizing photons $f_{\rm esc}$. To obtain the constraints on the fine-structure constant from the reionization epoch, there are two essential prerequisites. On the one hand, we should know the values of these two parameters with respect to $\alpha_{\text{EM}} =1/137.035999084(21)$. On the other hand, we should know the dependence of these two parameters on $\alpha/\alpha_{\rm EM}$, where $\alpha$ is the fine-structure constant in question. 

In this work, we make the fine-structure constant free during the whole ionization history and constrain its values with observational data in two models. In the $\Lambda$CDM$+\alpha/\alpha_{\text{EM}}$ model, the fine-structure constant remains a constant $\alpha$ at all redshifts. In the $\Lambda$CDM$+\alpha_{\text{rec}}/\alpha_{\text{EM}}+\alpha_{\text{rei}}/\alpha_{\text{EM}}$ model, there is a constant $\alpha_{\text{rec}}$ at the recombination epoch and another constant $\alpha_{\text{rei}}$ at the reionization epoch. In the latter model, we can do a consistency test of the fine-structure constant from the whole ionization history and have a glance at the redshift-varying fine-structure constant.

This paper is organized as follows.
In section~\ref{sec:ion}, the ionization history of our universe is reconsidered with the rescaled fine-structure constant. 
In subsection~\ref{sec:rec}, we introduce the recombination calculation related to $\alpha_{\text{rec}}$.
In subsection~\ref{sec:rei}, modifications of the reionization history are presented.
In section~\ref{sec:res}, we put constraints on free parameters in previous models with the same data combination including CMB measurement from Planck 2018~\cite{Planck:2018vyg,Planck:2019nip,Planck:2018lbu}, SFR density from UV data and IR data~\cite{Madau:2014bja}, $Q_\text{HII}$ constraints from observations of quasars~\cite{Bouwens:2015vha,Fan:2005es,Schenker:2014tda}, baryon acoustic oscillation (BAO) data~\cite{Beutler:2011hx,Ross:2014qpa,Alam:2016hwk,deSainteAgathe:2019voe,Ata:2017dya,Hou:2018yny}, as well as the measurements of Type Ia supernova~\cite{Scolnic:2017caz}. 
Finally, a brief summary is included in section~\ref{sec:sum}.

\section{Effects of the Fine-Structure Constant on Ionization History}
\label{sec:ion}
\begin{figure*}
\resizebox{215pt}{188pt}{\includegraphics{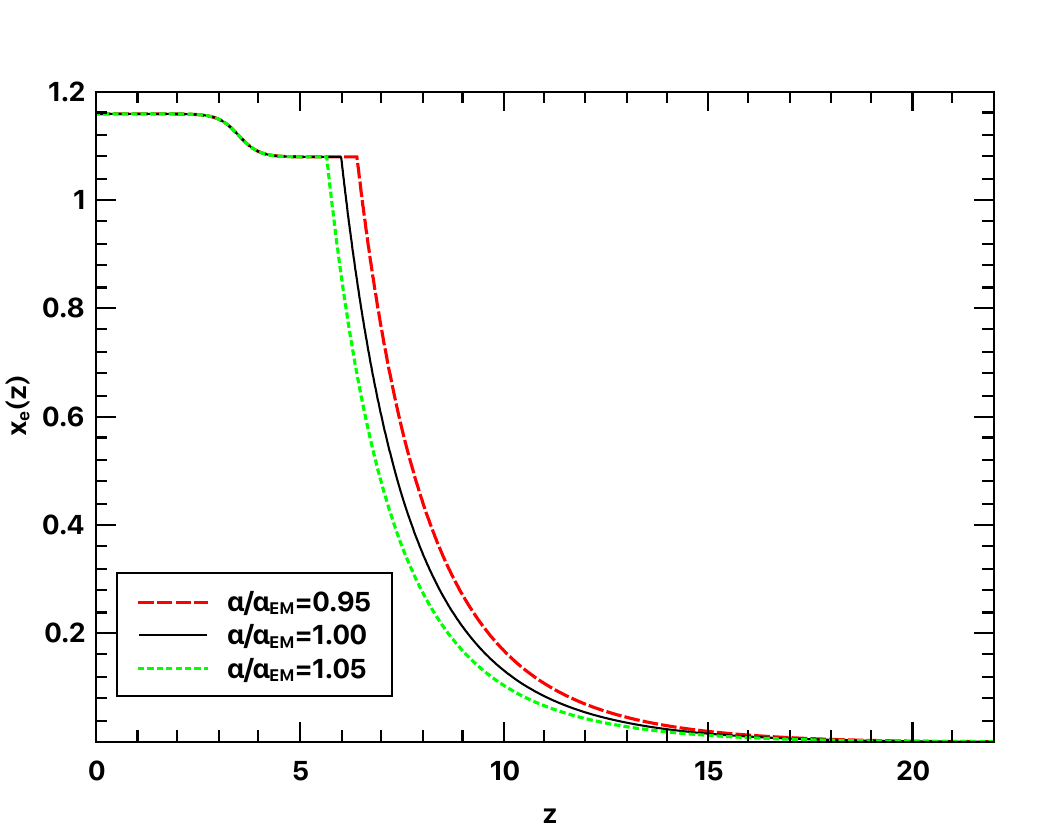}}
\resizebox{217pt}{188pt}{\includegraphics{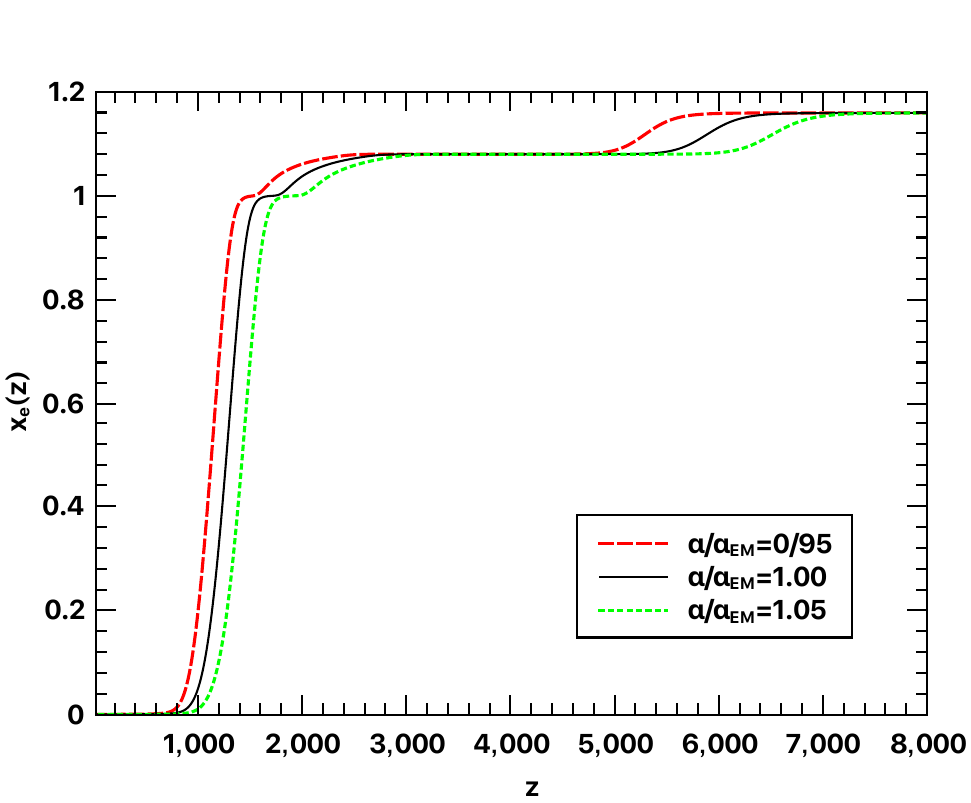}}
\caption{\label{fig:xe} The evolution of ionized fraction $x_e(z)$ during the reionization epoch (left) and the recombination epoch (right) in the $\Lambda$CDM$+\alpha/\alpha_{\text{EM}}$ model. 
For reionization, we assume the single ionization of Helium and the ionization of Hydrogen happened simultaneously.
The black solid curve indicates the fiducial case of $\alpha/\alpha_{\text{EM}}=1.00$. The red dashed curve corresponds to $\alpha/\alpha_{\text{EM}}=0.95$ and the green dotted one illustrates $\alpha/\alpha_{\text{EM}}=1.05$. Values of the other necessary parameters are set to their mean values in the $\Lambda$CDM$+\alpha/\alpha_{\text{EM}}$ model as Tab.~I shows. }
\end{figure*}

In this section, the effects of varying fine-structure constant on the ionization history 
are introduced and the evolution of ionized fraction is shown in Fig.~\ref{fig:xe}.   
Note we denote the fine-structure constant as $\alpha$  uniformly for both recombination epoch and reionization epoch.

  
\subsection{Effects of the Fine-Structure Constant on Recombination}
\label{sec:rec}
 
The evolution of ionized fraction $x_\text{e}=x_{\text{p}}+x_{\text{HeII}}$ is described by a set of differential equations during the recombination epoch~\cite{Planck:2014ylh}:
\begin{eqnarray}
\dfrac{dx_{\text{p}}}{dz} & =& \frac{C_{\text{H}}}{H_0(1+z)E(z)}\left[x_{\text{e}} x_{\text{p}} \langle n_{\text{H}}\rangle A_{\text{H}} -B_{\text{H}}(1-x_{\text{p}}) e^{-h \nu_{\text{H2s}}/kT_{\rm M}}\right], \\\label{e.a1}
\dfrac{dx_{\text{HeII} }}{dz}  & =& \frac{C_{\text{HeI}}}{H_0(1+z)E(z)}\left[x_{\text{e}} x_{\text{HeII}} \langle n_\text{H}\rangle A_\text{HeI}    -B_\text{HeI}(f_\text{He}-x_\text{HeII}) e^{-h \nu_{\text{HeI}, 2^1{\rm s}}/kT_{\rm M}}\right], \label{e.a2}\\
\dfrac{  dT_{\rm M}}{  dz} & =& \frac{8\sigma_{\rm T} a_{\rm R} T^4}{3 H_0 E(z) (1+z) m_{\rm e}c}(T_{\rm M}-T)+ \frac{2T_{\rm M}}{1+z}. \label{e.a3}
\end{eqnarray}
Here 
$A_i$ and $B_i$ ($i=\text{H, HeI}$) are the effective recombination rates and photoionization rates, respectively.
The coefficients $C_i$ are
\begin{eqnarray}
C_\text{H} &=& \frac{1+K_\text{H}\Lambda_\text{H} \langle n_\text{H}\rangle (1-x_\text{p})}{1+K_\text{H}(\Lambda_\text{H}+B_\text{H}) \langle n_\text{H}\rangle (1-x_\text{p})}\label{e.a4},\\
C_\text{HeI} &=& \frac{1+K_\text{HeI}\Lambda_\text{HeI} \langle n_\text{H}\rangle (f_\text{He}-x_\text{HeII})\,\hbox{e}^{h\nu_{\rm ps}/kT_{\rm M}}}{1+K_\text{HeI}(\Lambda_\text{He}+B_\text{HeI})\langle n_\text{H}\rangle (f_\text{He}-x_\text{HeII})\,\hbox{e}^{h\nu_{\rm ps}/kT_{\rm M}}},\label{e.a5}  
\end{eqnarray}
where the so-called ``$K$-quantities" $K_i$ represent the effective transition rates for the main resonances of hydrogen and helium. $\Lambda_i$ describe the two-photon decay rates.

As mentioned above, the fine-structure constant $\alpha$ implies the strength of electromagnetic interaction.
Therefore it affects the energy levels of hydrogen and helium atoms directly as $E_i\propto (\alpha/\alpha_{\text{EM}})^2$.
This leads to two main sequent changes.
Firstly, the transition frequencies $\nu_i$ are altered, such as the atomic transition rates of H 2s$\to$1s, He $2^1p\to1^1s$ and $2^3p\to1^1s$. 
Likewise, values of $\Lambda_i$, $A_i$, $B_i$, and $K_i$ are also rescaled. 
Secondly,
this leads to the effective temperature $T_{\text{eff}}\propto (\alpha/\alpha_{\text{EM}})^{-2}$ when we calculate $A_i$ and $B_i$. 
In addition, the Thomson scattering cross-section $\sigma_T$ ought to be modified.
Overall, above modifications can be summarized as following~\cite{Hart:2017ndk}:
\begin{equation}
\begin{split}
\Lambda_i \propto \left(\dfrac{\alpha}{\alpha_{\text{EM}}}\right)^8,\ \ 
A_i\propto \left(\dfrac{\alpha}{\alpha_{\text{EM}}}\right)^2,\ \  B_i\propto \left(\dfrac{\alpha}{\alpha_{\text{EM}}}\right)^5, \ \  K_i \propto \left(\dfrac{\alpha}{\alpha_{\text{EM}}}\right)^{-6},\ \  T_{\text{eff}}\propto \left(\dfrac{\alpha}{\alpha_{\text{EM}}}\right)^{-2}, \ \ \sigma_T \propto \left(\dfrac{\alpha}{\alpha_{\text{EM}}}\right)^{2}.
\end{split}
\end{equation}
As the right figure of Fig.~\ref{fig:xe} shows, recombination begins earlier with higher $\alpha$.

\subsection{Effects of the Fine-Structure Constant on Reionization}
\label{sec:rei}

Previous researches reveal stars and galaxies contributed the majority of ionizing photons during the epoch of reionization~\cite{Madau:2014bja,Robertson:2015uda,Paoletti:2021gzr,Hazra:2019wdn,Qin:2020xrg,Mitra:2011uv,Mitra:2016olz,Gorce:2017glg}.
In this work, we reconstruct the reionization history with star formation history, instead of the instantaneous reionization model.

The Thomson optical depth is given by
\m
\tau(z) = c  \sigma_{\text{T}} \int_0^z (1+f_{\text{He}}) Q_{\text{HII}}(z')n_{\text{H}}(z') (1+z')^{-1} H^{-1}(z')dz'.
\n
Here $c$ is the speed of light and $H(z)$ is the redshift-dependent Hubble parameter. The ionization fraction $f_{\rm He}$ is $0.08$ (or $0.16$) for the singly (or fully) ionized Helium.
The ionized fraction $Q_{\text{HII}}$ (equivalent to $x_\text{p}$ physically) can be calculated by the following differential equation~\cite{Robertson:2015uda}:
\m
\label{eq:q}
\dot{Q}_{\text{HII}} =  -\dfrac{Q_{\text{HII}}}{t_{\text{rec}}} + \dfrac{\dot{n}_{\text{ion}}}{\langle n_\text{H}\rangle} .
\n
Here $t_{\text{rec}}$ is the recombination time defined by~\cite{Robertson:2015uda,Bouwens:2015vha}
\m
t_{\text{rec}}&=&\left[ C_{\text{HII}} A_{\text{H}} \left(1+\dfrac{Y_\text{p}}{4x_{\text{p}}}\right) \langle n_\text{H}\rangle (1+z)^3\right]^{-1}\nonumber\\
 & = & 0.88\  \text{Gyr} \(\dfrac{1+z}{7}\)^{-3} \(\dfrac{T_0}{2\times 10^4 \text{K}}\)^{-0.7} \( \dfrac{C_{\text{HII}}}{3}\)^{-1}  ,
\n
where
$T_0 \sim 2\times 10^4$ K is the temperature of the ionizing hydrogen gas and 
$C_{\text{HII}} \sim 3$ is the clumping factor of ionized hydrogen.
The second term of Eq.~\ref{eq:q} represents the effect of SFR.
The growth rate of number density for ionized photons is~\cite{Robertson:2015uda}
\m
\label{eq:sfr}
\dot{n}_\text{ion} = f_\text{esc} \xi_\text{ion} \rho_\text{SFR},
\n
where $f_\text{esc}$ is the escaping fraction, which indicates the fraction of
the number of escaping Lyman continuum photons to that of Lyman continuum
photons produced in a galaxy.
$\xi_\text{ion}$ transforms a UV luminosity density into the Lyman continuum photon emission rate of a star-forming galaxy.
We adopt the widely used values $f_\text{esc}=0.2$~\cite{Robertson:2013bq} and $\log_{10} \xi_\text{ion} = 53.14\  [\text{Lyc} \cdot \text{photons} \cdot  \text{s}^{-1}\cdot  \text{M}_{\odot}^{-1}\cdot  \text{yr}]$~\cite{Dunlop:2012ym}.
$\rho_{\text{SFR}}$ is the SFR density, which reads in units of $\text{M}_{\odot} \cdot \text{yr}^{-1} \cdot \text{Mpc}^{-3}$ as follows~\cite{Robertson:2015uda,Chen:2021gwj}:
\m
\label{eq:rhosfr}
\rho_{\text{SFR}} = \dfrac{a_{\text{p}} (1+z)^{b_{\text{p}}}}{1+[(1+z)/c_{\text{p}}]^{d_{\text{p}}}} - \dfrac{a_{\text{p}} (1+22)^{b_{\text{p}}}}{1+[(1+22)/c_{\text{p}}]^{d_{\text{p}}}}.
\n
$a_\text{p}$, $b_\text{p}$, $c_\text{p}$ and $d_\text{p}$ are four undetermined coefficients.
We add the second term because there is almost no 21 cm absorption signal before $z \sim 22$ in the 21 cm absorption profiles~\cite{Bowman:2018yin}.

Considering a varying fine-structure constant in star formation, the recombination time $t_{\text{rec}}$ should be rescaled as $t_{\text{rec}} \propto A_{\text{H}}^{-1} \propto (\alpha/\alpha_{\text{EM}})^{-2}$.
The escaping rate of ionizing photons $f_\text{esc}$ would be inversely propotional to the Thomson cross-section, which implies $f_\text{esc} \propto \sigma_T^{-1} \propto (\alpha/\alpha_{\text{EM}})^{-2}$.
The ionizing photon production rate $\xi_\text{ion}$ would be rescaled as $\xi_\text{ion} \propto (\alpha/\alpha_{\text{EM}})^{-2}$ because the Lyman limits shift shortward by $(\alpha/\alpha_{\text{EM}})^2$. 
We show the ionized fraction during reionization epoch in the left picture of Fig.~\ref{fig:xe} for different $\alpha/\alpha_{\text{EM}}$.

\section{Results}
\label{sec:res} 

In this section, we put constraints on the fine-structure constant and other cosmological parameters by reconstructing the ionization history of our universe from observational data.
We apply the following data combination: the latest measurements of CMB power spectra---Planck 2018 TT,TE,EE$+$lowE$+$lensing~\cite{Planck:2018vyg,Planck:2019nip,Planck:2018lbu}, the SNIa Pantheon sample~\cite{Scolnic:2017caz}, the BAO measurements at redshifts $z = 0.106,0.15,0.32,0.57,1.52,2.34$ (named 6dFGS~\cite{Beutler:2011hx}, MGS~\cite{Ross:2014qpa}, DR12~\cite{Alam:2016hwk}, DR14~\cite{deSainteAgathe:2019voe,Ata:2017dya,Hou:2018yny} respectively), the SFR density from UV and IR data~\cite{Madau:2014bja},  as well as the $Q_\text{HII}$ constraints between $5.0\leq z \leq 8.0$ from observations of Gunn-Peterson optical depth and Ly-$\alpha$ emission in galaxies~\cite{Bouwens:2015vha,Fan:2005es,Schenker:2014tda}.
In this work, we use the SFR data below $z\sim 8$ to reconstruct the reionization history. Certainly, some researches, such as Ref.~\cite{Bouwens:2015vha,Ishigaki_2018}, have provided the SFR densities extending to $z\sim 10$. However, the data at high-$z$ vary dramatically with different truncation magnitudes $M_{\text{trunc}}$ and their errors also get larger with higher redshifts. So, our choice would not impact the results significantly.
We modify CAMB~\cite{Lewis:1999bs} and CosmoRec~\cite{Shaw:2011ez} packages to calculate the optical depth and CMB power spectra. Then the Markov Chain Monte Carlo sampler---CosmoMC~\cite{Lewis:2002ah,Lewis:2013hha} is applied to constrain the free parameters.
In the first model, named the $\Lambda$CDM+$\alpha/\alpha_{\text{EM}}$ model thereafter, there are ten free parameters: $\{\Omega_{\text{b}}h^2, \Omega_{\text{c}}h^2, 100\theta_{\text{MC}}, \text{ln} \(10^{10} A_\text{s}\), n_{\text{s}}, a_\text{p}, b_\text{p}, c_\text{p}, d_\text{p} ,\alpha/\alpha_{\text{EM}} \}$. Five of them are basic parameters in the $\Lambda$CDM model: $\Omega_\text{b}h^2$ and $\Omega_\text{c}h^2$ are the density of baryons and cold dark matter today respectively, $100\theta_\text{MC}$ is 100 times the ratio of angular diameter distance to the large scale structure sound horizon, $A_\text{s}$ is the amplitude of the power spectrum of primordial curvature perturbations, and $n_\text{s}$ is the scalar spectrum index. Four of them, $a_\text{p}, b_\text{p}, c_\text{p},d_\text{p}$ are parameters of SFR density as described in section~\ref{sec:rei}. Lastly, $\alpha/\alpha_{\text{EM}}$ indicates the ratio of fine-structure constant $\alpha$ to its fiducial value $\alpha_{\text{EM}}$. It remains unchanged during the whole ionization history.
In the second model, named the $\Lambda$CDM$+$ $\alpha_{\text{rec}}/\alpha_{\text{EM}}$$+$ $\alpha_{\text{rei}}/\alpha_{\text{EM}}$ model, there are eleven free parameters: $\{\Omega_{\text{b}}h^2, \Omega_{\text{c}}h^2, 100\theta_{\text{MC}}, \text{ln} \(10^{10} A_\text{s}\), n_{\text{s}}, a_\text{p}, b_\text{p}, c_\text{p}, d_\text{p} ,\alpha_{\text{rec}}/\alpha_{\text{EM}}, \alpha_{\text{rei}}/\alpha_{\text{EM}} \}$. We use $\alpha_{\text{rec}}$ and $\alpha_{\text{rei}}$ to denote the fine-structure constant during recombination and reionization epoch respectively.

\begin{figure*}
\resizebox{215pt}{188pt}{\includegraphics{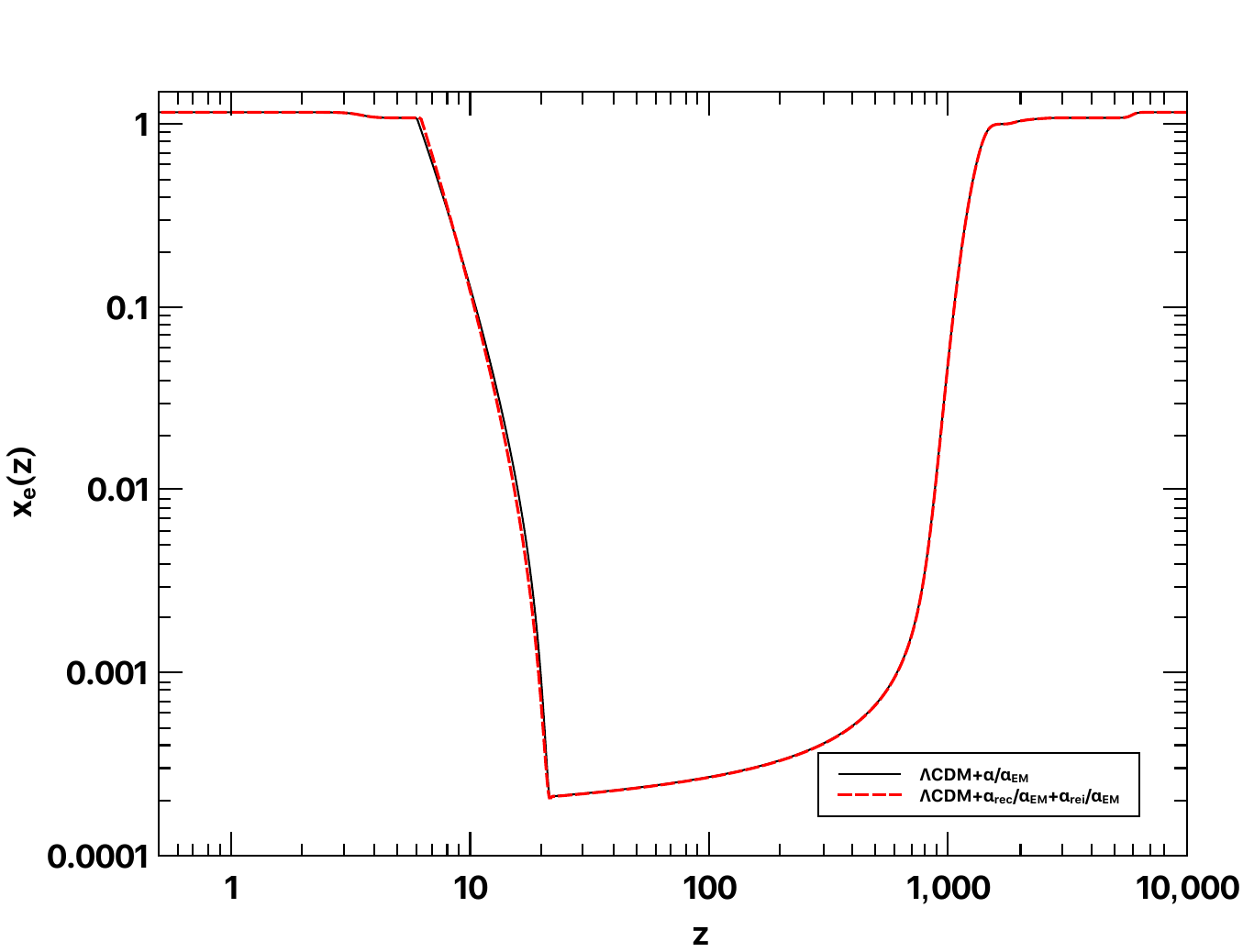}}
\resizebox{215pt}{189pt}{\includegraphics{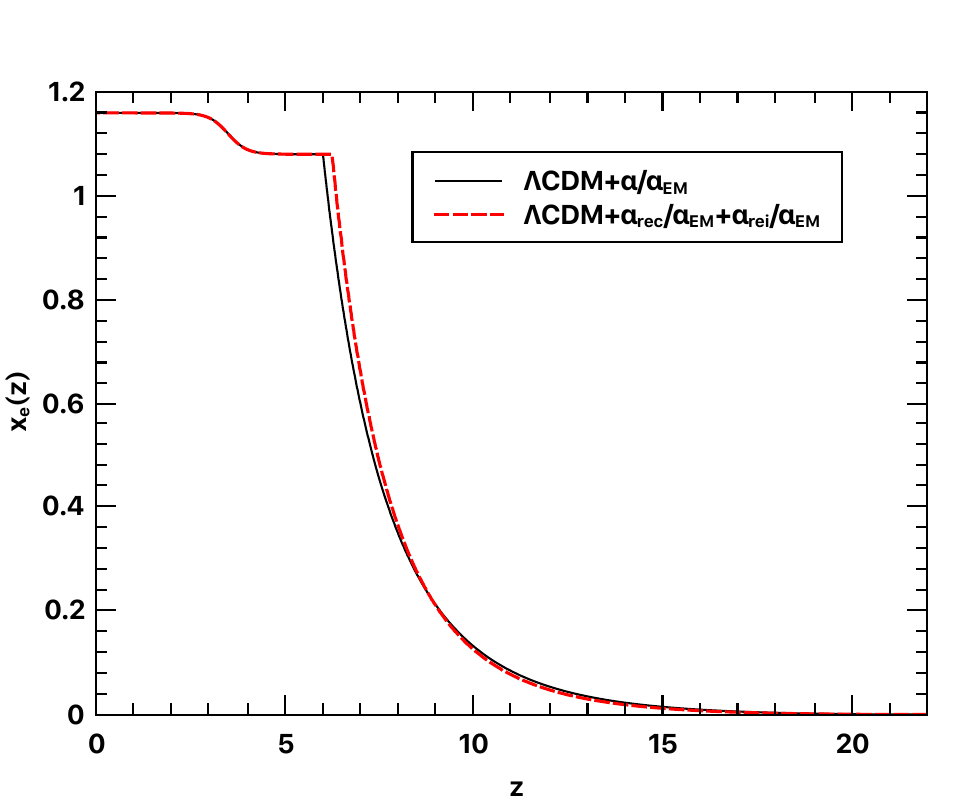}}
\caption{\label{fig:xe-real} The left picture shows the evolution of ionized fraction $x_e(z)$ in the $\Lambda$CDM$+\alpha/\alpha_{\text{EM}}$ model and the $\Lambda$CDM$+\alpha_{\rm rec}/\alpha_{\text{EM}}+\alpha_{\rm rei}/\alpha_{\text{EM}}$ model. The right one shows the reionization histories. The black solid curve corresponds to the $\Lambda$CDM$+\alpha/\alpha_{\text{EM}}$ model and the red dashed one illustrates the $\Lambda$CDM$+\alpha_{\rm rec}/\alpha_{\text{EM}}+\alpha_{\rm rei}/\alpha_{\text{EM}}$ model. Values of all the parameters are set to their mean values as Tab.~I shows. }
\end{figure*}

Our results are summarized in Tab.~I. We have listed the 68$\%$ limits of the free parameters and necessary derived parameters in these two models.
The ratio of fine-structure constant reads $\alpha/\alpha_{\text{EM}}=1.000042^{+0.001983}_{-0.001972}$ at 68$\%$C.L. in the $\Lambda$CDM+$\alpha/\alpha_{\text{EM}}$ model, which is well consistent with $1$. 
The uncertainties decrease slightly than previous works which modify the fine-structure constant at only the recombination epoch. For example, the Planck Collaboration released  $\Delta \alpha/\alpha_{\rm EM}=(3.6\pm3.7)\times 10^{-3}$ at 68$\%$ C.L.~\cite{Planck:2014ylh} and Ref.~\cite{Hart:2017ndk} obtained $\alpha/\alpha_{\rm EM}=0.9993\pm 0.0025$ at 68$\%$ C.L..  
Comparing other measurements from astrophysics, our constraints are tighter than $\Delta \alpha/\alpha_{\rm EM} < 10^{-2} -10^{-3}$ from the abundance of light elements during BBN~\cite{Mosquera:2013dga}, but weaker than constraints of $\Delta \alpha/\alpha_{\rm EM} \sim 10^{-7}-10^{-8}$ from the 1.8 billion-year-old natural nuclear reactor at the Oklo Uranium Mine in Gabon~\cite{Damour:1996zw}, $\Delta \alpha/\alpha_{\rm EM} \sim 10^{-5}-10^{-6}$ from the spectral lines of quasars~\cite{Murphy:2001nu,Murphy:2000pz,Murphy:2000ns} and so on~\cite{Liu:2021mfk}.
However, in the $\Lambda$CDM+$\alpha_{\text{rec}}/\alpha_{\text{EM}}+\alpha_{\text{rei}}/\alpha_{\text{EM}}$ model, our results show $\alpha_{\text{rec}}/\alpha_{\text{EM}}=1.001494^{+0.002041}_{-0.002063}$ and $\alpha_{\text{rei}}/\alpha_{\text{EM}}=0.8540342^{+0.031678}_{-0.027209}$ at 68$\%$C.L..
It indicates $\alpha_{\text{rec}}$ is inconsistent with $\alpha_{\text{rei}}$ by 4.64$\sigma$.
The evolution of ionized fraction $x_e(z)$ in these two models is shown in Fig.~\ref{fig:xe-real}. They coincide with each other almostly during the recombination epoch, but differ slightly at the reionization epoch. We also provide a separate figure to show the small difference. 
The straightforward explanation for this inconsistency is that the fine-structure constant may vary with redshift.
However, the value at low redshift is much smaller than $\alpha_{\text{EM}}$. 
Therefore, we prefer another conservative explanation that the observational data about SFR we used are not good enough and a calibration of some parameters about SFR is necessary. Since $f_{\rm esc}$ and $\xi_{\rm ion}$ appear in product form, for example, above inconsistency can be solved by rescaling this product to $f_{\rm esc}\xi_{\rm ion}\left(\frac{\alpha_{\rm rei}}{\alpha_{\rm EM}}\right)^{-4}$ if all the difference between $\alpha_{\text{rei}}$ and $\alpha_{\text{EM}}$ is due to the uncertainty in the product of $f_{\rm esc}$ and $\xi_{\rm ion}$.
 In fact, this calibration is acceptable. Because the reasonable constraints on $f_{\rm esc}$ and $\xi_{\rm ion}$ at the epoch of reionization have not been obtained so far~\cite{Ishigaki_2018}.

\begin{figure}[]
\begin{center}
\includegraphics[scale=0.2]{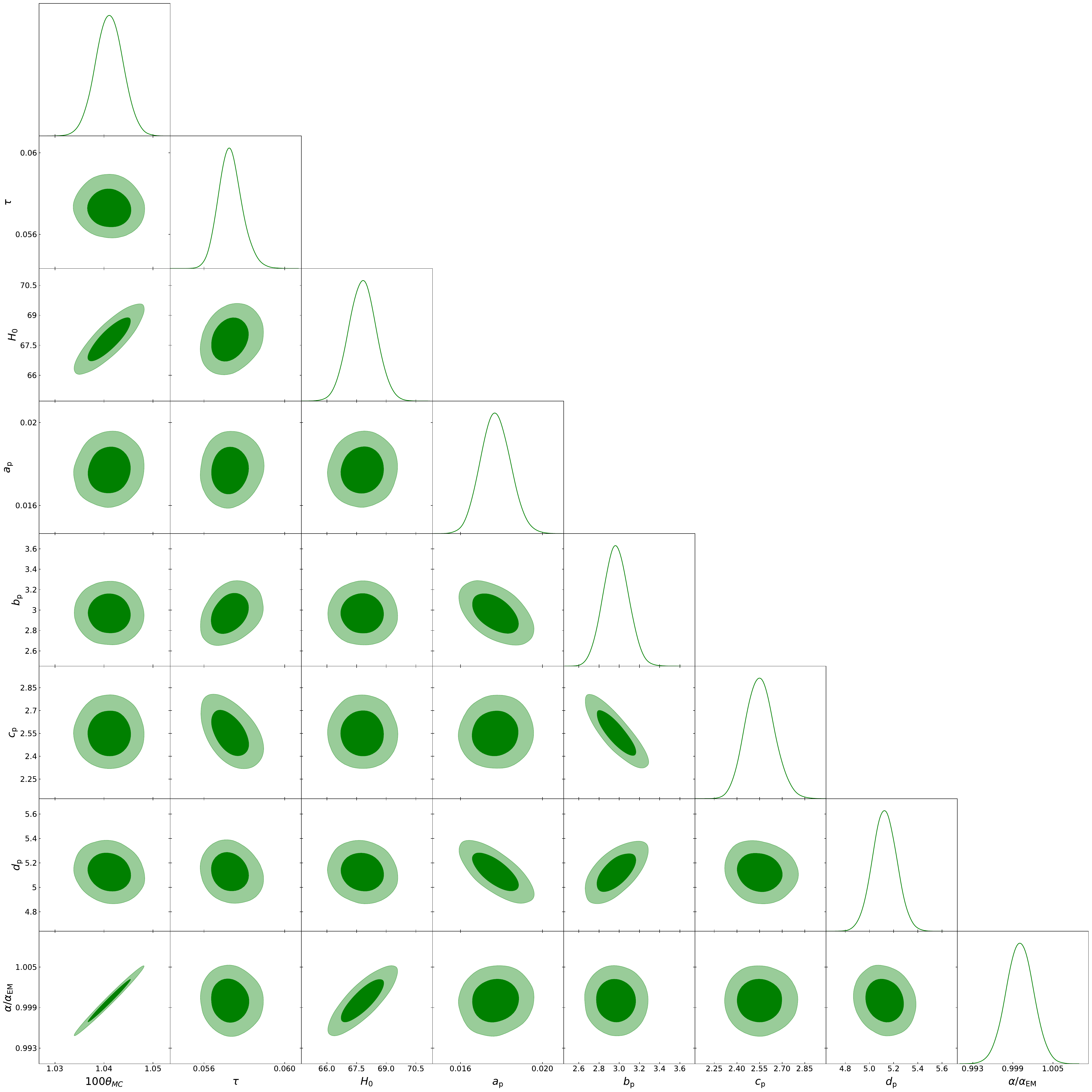}
\end{center}
\caption{The triangular plot of $100\theta_{\text{MC}},\tau,H_0,a_{\rm p},b_{\rm p},c_{\rm p},d_{\rm p},\alpha/\alpha_{\text{EM}}$ in the $\Lambda$CDM$+\alpha/\alpha_{\text{EM}}$ model from observational data.}
\label{fig:alpha1}
\end{figure}

\begin{figure}[]
\begin{center}
\includegraphics[scale=0.19]{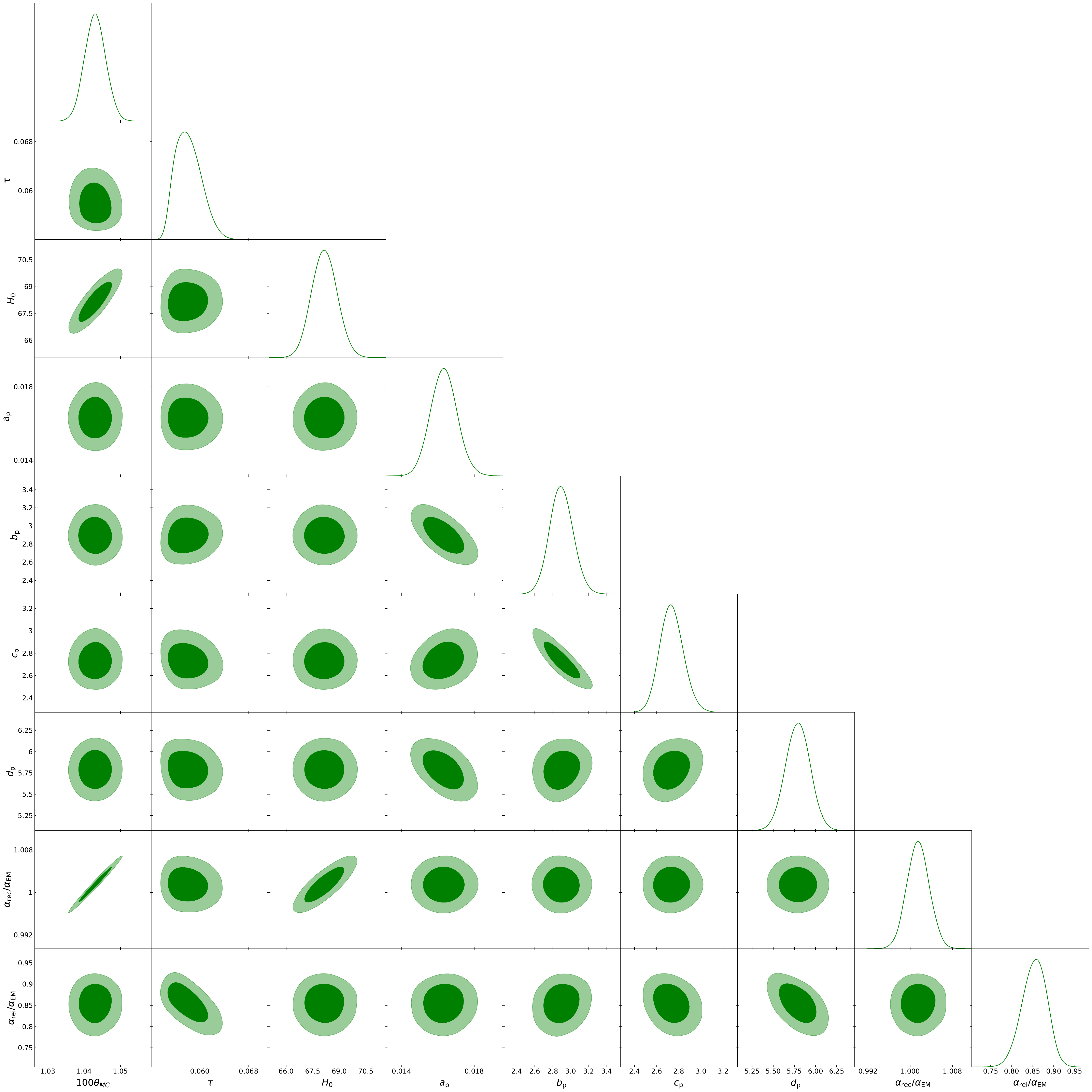}
\end{center}
\caption{The triangular plot of $100\theta_{\text{MC}},\tau,H_0,a_{\rm p},b_{\rm p},c_{\rm p},d_{\rm p},\alpha_{\text{rec}}/\alpha_{\text{EM}},\alpha_{\text{rei}}/\alpha_{\text{EM}}$ in the $\Lambda$CDM$+\alpha_{\text{rec}}/\alpha_{\text{EM}}+\alpha_{\text{rec}}/\alpha_{\text{EM}}$ model from observational data.}
\label{fig:alpha2}
\end{figure}

Moreover, we find the optical depth $\tau=0.0573_{-0.0006}^{+0.0005}$ at 68$\%$C.L. in the $\Lambda$CDM+$\alpha/\alpha_{\text{EM}}$ model and  $\tau=0.0559_{-0.0076}^{+0.0069}$ in the $\Lambda$CDM+$\alpha_{\text{rec}}/\alpha_{\text{EM}}+\alpha_{\text{rei}}/\alpha_{\text{EM}}$ model. These little larger values derived by integrating from $z=0$ to $z=22$ are still consistent with $\tau=0.054\pm0.007$ in the instantaneous reionization model given by Planck 2018.
Also, we list values of Hubble constant, $H_0=67.80\pm0.67$ km $\cdot$ s$^{-1}$ $\cdot$ Mpc$^{-1}$ in the $\Lambda$CDM+$\alpha/\alpha_{\text{EM}}$ model and $H_0=68.17\pm 0.70$ km $\cdot$ s$^{-1}$ $\cdot$ Mpc$^{-1}$ in the $\Lambda$CDM+$\alpha_{\text{rec}}/\alpha_{\text{EM}}+\alpha_{\text{rei}}/\alpha_{\text{EM}}$ model.
Therefore, the varying fine-structure constant can relieve the so-called Hubble tension slightly. 
Furthermore, we show the triangular plots of several parameters in previous two models in Fig.~\ref{fig:alpha1} and Fig.~\ref{fig:alpha2}. We find that there is an expected strong degeneracy between $100\theta_{\rm MC}$ and the fine-structure constant during the recombination as well as an expected degeneracy between $\tau$ and the fine-structure constant during the reionization.

\begin{table}
\label{tb:result}
\caption{The 68$\%$ limits for the cosmological parameters in the $\Lambda$CDM+$\alpha/\alpha_{\text{EM}}$ and $\Lambda$CDM$+$ $\alpha_{\text{rec}}/\alpha_{\text{EM}}$$+$ $\alpha_{\text{rei}}/\alpha_{\text{EM}}$ model. }
\begin{tabular}{p{3.5 cm}<{\centering}|p{5cm}<{\centering} p{5cm}<{\centering}   }
\hline
                  & $\Lambda$CDM+$\alpha/\alpha_{\text{EM}}$                   & $\Lambda$CDM$+$ $\alpha_{\text{rec}}/\alpha_{\text{EM}}$$+$ $\alpha_{\text{rei}}/\alpha_{\text{EM}}$ \\
\hline
$\Omega_\text{b} h^{2}$& $0.02244\pm 0.00014$        & $0.02241\pm0.00014$ \\
$\Omega_\text{c} h^{2}$& $0.1191\pm 0.0013$          & $0.1198\pm 0.0013$\\
$100\theta_\text{MC}$&$1.04107\pm0.00272$&$1.04305_{-0.00284}^{+0.00280}$ \\
$\ln(10^{10}\text{A}_\text{s})$&$3.0484\pm0.0057$& $3.0492_{-0.0068}^{+0.0064}$\\
$n_s$             & $0.9670_{-0.0064}^{+0.0063}$               & $0.9630\pm 0.0064$ \\
\hline
$a_{\text{p}}$[$\text{M}_{\odot} \cdot \text{yr}^{-1} \cdot \text{Mpc}^{-3}$]   & $0.01772_{-0.00071}^{+0.00070}$ &$0.01633_{-0.00071}^{+0.00070}$\\
 $b_{\text{p}}$   &   $2.968\pm0.121 $              & $2.897_{-0.134}^{+0.124} $\\
 $c_{\text{p}}$   &   $2.554_{-0.101}^{+0.088}$     &  $2.736_{-0.112}^{+0.098}$  \\
 $d_{\text{p}}$   & $5.125\pm0.097$       & $5.791\pm0.140$ \\
 \hline
 $\alpha/\alpha_{\text{EM}}$ & $1.000042_{-0.001972}^{+0.001983}$ &-\\
 $\alpha_{\text{rec}}/\alpha_{\text{EM}}$ & - &$1.001494_{-0.002063}^{+0.002041}$\\
 $\alpha_{\text{rei}}/\alpha_{\text{EM}}$ & - &$0.854034_{-0.027209}^{+0.031678}$\\
 \hline
 $\tau$           &   $0.0573_{-0.0006}^{+0.0005} $ &$0.0559_{-0.0076}^{+0.0069}$      \\
 $10^{4}Q_\text{HII}(z=22)$&$2.133\pm0.019$   &$2.125\pm0.020$ \\
$H_0$[km $\cdot$ s$^{-1}$ $\cdot$ Mpc$^{-1}$]&$67.80\pm0.67$&$68.17\pm 0.70$  \\
 \hline
\end{tabular}
\end{table}

\section{Summary and discussion}
\label{sec:sum}

In this paper, we investigate how the fine-structure constant $\alpha$ affect the whole ionization history of our universe, including both the recombination and reionization epoch. 
$\alpha$ influences the Thomson scattering cross-section and atomic energy levels, then modify the ionization history, which are reflected in the CMB power spectra.
Therefore we constrain the fine-structure constant and other free parameters in two $\alpha$-related models
from data combination of Planck 2018, BAO data, PANTHON samples, SFR density from UV and IR measurements, and the $Q_\text{HII}$ constraints, using CosmoRec$+$CAMB$+$CosmoMC packages.
Comparing previous studies on constraining $\alpha$ from CMB, we take the influence of  $\alpha$ on the reionization epoch into consideration and reconstruct the reionization history with SFR density.
In the $\Lambda$CDM$+\alpha/\alpha_{\text{EM}}$ model, the fine-structure constant behaves as a constant and we find its value is well consistent with the standard one.
In the $\Lambda$CDM$+\alpha_{\text{rec}}/\alpha_{\text{EM}}+\alpha_{\text{rei}}/\alpha_{\text{EM}}$ model, the fine-structure constant reads $\alpha_{\text{rec}}/\alpha_{\text{EM}}=1.001494^{+0.002041}_{-0.002063}$ at 68$\%$ C.L. during the recombination epoch and $\alpha_{\text{rei}}/\alpha_{\text{EM}}=0.854034^{+0.031678}_{-0.027209}$ at the reionization epoch.
The value of $\alpha_{\text{rei}}$ deviates from $\alpha_{\text{rec}}$ by 4.64$\sigma$.

If above such an inconsistency results from the uncertainty in the product of $f_{\rm esc}$ and $\xi_{\rm ion}$, more refined calibration is necessary.  
That is to say, instead of the typical values of these two parameters  $f_\text{esc}=0.2$ and $\log_{10} \xi_\text{ion} = 53.14\  [\text{Lyc} \cdot \text{photons} \cdot  \text{s}^{-1}\cdot  \text{M}_{\odot}^{-1}\cdot  \text{yr}]$, we should turn to their improved counterparts.
Given enough star-forming galaxies at different redshifts, we can use the observed H$\alpha$ and UV-continuum fluxes of them to improve the constraints on $\log_{10} \xi_\text{ion}$~\cite{Bouwens:2016} and use the direct detection of Lyman-continuum emission from them to improve constraints on $f_{\rm esc}$~\cite{Steidel:2001,Shapley:2006fp}.
Also a better fitting formula and measurements with a higher precision of SFR density would help us reconstruct the reionization history and acquire a more precise value of $\alpha$. It would be interesting to evaluate how a new fitting formula including the steeper slopes of $\log \rho_{\text{SFR}}$ at $z \sim 8-10$~\cite{Ishigaki_2018} affects the results.
Moreover, the fine-structure constant may be a redshift-dependent variation which is beyond the simple models of $\alpha$ assumed in the present paper.
We expect the problem of fine-structure constant to obtain a much more robust description with future researches using future high precision data.

\vspace{5mm}
\noindent {\bf Acknowledgments}

We acknowledge the use of HPC Cluster of Tianhe II in National Supercomputing Center in Guangzhou. Ke Wang is supported by grants from NSFC (grant No. 12005084) and grants from the China Manned Space Project with NO. CMS-CSST-2021-B01. Lu Chen is supported by grants from NSFC (grant No. 12105164). This work has also received funding from project ZR2021QA021 supported by Shandong Provincial Natural Science Foundation.



\end{document}